\documentstyle{article}
\input{tcilatex}

\begin{document}

Title: Remarks on the correspondence of the relativity and causality
principles

Author: Alexander L. Kholmetskii

Comments: 4 pages

Subj-class: General Physics

Journal-ref: Apeiron 8, No 1, (2001)

\bigskip

\qquad A particular problem about special kind of two light pulses
propagation has been considered in cases of inertial motion, constant
homogeneous gravitation field and progressive non-inertial motion with
constant acceleration. A contradiction between the causality principle and
relativity theory has been revealed.

\bigskip

{\bf 1. Introduction}

It is well-known that two Einstein's postulates form a basis of special
theory of relativity (STR):

1. All inertial reference frames are equivalent to each other.

2. A light velocity in vacuum does not depend on a velocity of emitter.

At present time one can see some restrictions on application of these
postulates. For example, the second postulate formally is not valid in
arbitrary co-ordinates. In order to overcome the restrictions, ref. [1]
proposes a relativistic postulate in the following short form: a geometry of
empty physical space-time is pseudo-Euclidean. Indeed, such a postulate
allows to operate with any arbi-trary co-ordinates of inertial reference
frames and additionally includes into STR a case of non-inertial motion. The
latter statement follows from the obvious fact that any arbitrary motion
does not influence on a properties of geometry of physical space-time, it
continues to be pseudo-Euclidean. (From a formal point of view it means that
the curvature tensor is equal to zero in both inertial and non-inertial
frames). According to the author's opinion, such a formulation of
relativistic postulate allows to deeper understand a physical essence of
relativity theory. For example, it allows to advance a problem about
correspondence of STR to the causality principle. At the first sight, this
problem seems to be trivial due to a finiteness of light velocity in STR.
Nevertheless, the present paper finds some additional questions in this
topic in the following particular physical problem.

\bigskip

{\bf 2. A special case of two light pulses propagation: in inertial
reference frame, in constant homogeneous gravitation field, and in rigid
non-inertial frame}

\bigskip

{\bf 2.1. Case of inertial motion}

Let us consider the following problem in Cartezian inertial reference frame.

\qquad Let a short light pulse be emitted from the point $x=0$ along the
axis $x$. Let a number of re-emitters of light RL$_{m}$ be located along the 
$x$-axis in some points $x_{m}$ (RL$_{0}$ is located in the point $x=0$ and,
for simplicity, all $\Delta x_{m}=x_{m+1}-x_{m}$ are equal to each other).
When a light pulse arrives at each re-emitter, it is absorbed by it, and
after a fixed interval of its own time $\Delta \tau _{0\text{ }}$is emitted
by RL along the $x$-axis again.

\qquad Further, let the second light pulse be emitted from the point $x=0$
at such moment of time (taken as $t=0$), when the first light pulse has a
coordinate

$0<\Delta x\leq x_{1\text{.}}\qquad \qquad \qquad \qquad \qquad \qquad
\qquad \qquad \qquad $\qquad \qquad \qquad \qquad \qquad \qquad \qquad
\qquad (1)

One requires to find the times $t_{1}$ and $t_{2}$, where $t_{1}$ is the
moment of time when the first (right) light pulse is emitted by RL$_{n}$,
while $t_{2}$ is the moment of time when the second (left) pulse is reaching
RL$_{n}$, and $n$ is some number.

\qquad Due to the condition (1), the most general expression for $t_{1}$\
can be written as

$t_{1}=t_{x_{1}-\Delta x}+\sum_{m=1}^{n-1}t_{m}+\sum_{m=1}^{n}\Delta t_{m}$,$%
\qquad $\qquad \qquad \qquad \qquad \qquad \qquad \qquad \qquad \qquad
\qquad \qquad (2)

where $t_{x_{1}-\Delta x}$ is the propagation time of the first (right)
light pulse from the point $\Delta x$ to point $x_{1}$, $t_{m}$ is the
propagation time of the right pulse from RL$_{m}$ to RL$_{m+1}$, and $\Delta
t_{m}$\ is the time interval $\Delta \tau _{0}$\ for RL$_{m}$, remitting the
right pulse. The general expression for $t_{2}$ is:

$t_{2}=t_{x_{1}-0}^{\prime }+\sum_{m=1}^{n-1}t_{m}^{\prime
}+\sum_{m=0}^{n-1}\Delta t_{m}^{\prime }$,\qquad \qquad \qquad \qquad \qquad
\qquad \qquad \qquad \qquad \qquad \qquad \qquad (3)

where $t_{x_{1}-0}^{\prime }$ is the propagation time of the second (left)
light pulse from the point $\Delta x$\ to point $x_{1}$, $t_{m}^{\prime }$\
is the propagation time of the left pulse from the RL$_{m}$ to RL$_{m+1}$,
and $\Delta t_{m}^{\prime }$\ is the time interval $\Delta \tau _{0}$\ for RL%
$_{m}$, emitting the left pulse. From (3) and (2)

$t_{2}-t_{1}=\Delta t+\sum\limits_{m=1}^{n-1}\left( t_{m}^{\prime
}-t_{m}\right) +\left( \sum\limits_{m=0}^{n-1}\Delta t_{m}^{\prime
}-\sum\limits_{m=1}^{n}\Delta t_{m}\right) $,\qquad \qquad \qquad \qquad
\qquad \qquad (4)

where $\Delta t=t_{x_{1}-0}^{\prime }-t_{x_{1}-\Delta x}$.

\qquad Now let us ask the question: is it possible to implement the equality 
$t_{2}-t_{1}=0$? It is obvious, such an equality would mean an absolute
event: a meeting of both light pulses considered in the spatial point $x_{m}$%
.

\qquad This problem has a trivial solution in inertial reference frame. Here 
$t_{m}=t_{m}^{\prime }$, $\Delta t_{m}=\Delta t_{m}^{\prime }$, and all $%
\Delta t_{m}$ are equal to each other for any $m$. Hence, $%
t_{2}-t_{1}=\Delta t$, and the equality of $t_{1}$ and $t_{2}$ is
impossible. This result means that at the moment of time when the second
(left) pulse is reaching RL$_{n}$ in $x_{n}$ point, the first (right) pulse
already has a space coordinate $x_{n}+\Delta x$.

\qquad \bigskip

{\bf 2.2. Case of constant homogeneous gravitation field}

In this case

$t_{m}=t_{m}^{\prime }\left( 1+\frac{\varphi _{m}}{c^{2}}\right) $\qquad
\qquad \qquad \qquad \qquad \qquad \qquad \qquad \qquad \qquad \qquad \qquad
\qquad \qquad \qquad (5)

due to independence of metric tensor on time co-ordinate, and

$\Delta \tau _{0}\approx \Delta t_{m}$\qquad \qquad \qquad \qquad \qquad
\qquad \qquad \qquad \qquad \qquad \qquad \qquad \qquad \qquad \qquad \qquad
(6)

in the approximation of weak gravitation field. Here $c$ is the light
velocity in vacuum, and $\varphi $\ is gravitation potential. (Further we
take $\varphi _{0}=0$).

\qquad One follows from (6) that all values $\Delta t_{m}$\ and $\Delta
t_{m}^{\prime }$\ are equal to each other, and

$\sum\limits_{m=0}^{n-1}\Delta t_{m}^{\prime }-$.$\sum\limits_{m=1}^{n}%
\Delta t_{m}=\Delta t_{0}^{\prime }+\sum\limits_{m=1}^{n-1}\Delta
t_{m}^{\prime }-\sum\limits_{m=1}^{n-1}\Delta t_{m}-\Delta t_{n}=\Delta
t_{0}-\Delta t_{n}=$

=$\frac{\Delta \tau _{0}\left( \varphi _{n}/c^{2}\right) }{1+\left( \varphi
_{n}/c^{2}\right) }\approx \Delta \tau _{0}\frac{\varphi _{n}}{c^{2}}$\qquad
\qquad \qquad \qquad \qquad \qquad \qquad \qquad \qquad \qquad \qquad \qquad
\qquad \qquad (7)

in this approximation. Substituting (7) and (5) into (4), one gets:

$t_{2}-t_{1}=\Delta t+\Delta \tau _{0}\frac{\varphi _{n}}{c^{2}}$.\qquad
\qquad \qquad \qquad \qquad \qquad \qquad \qquad \qquad \qquad \qquad \qquad
\qquad \qquad (8)

Hence, the equality of $t_{1}$ and $t_{2}$ is implemented under the condition

$\Delta t=-\Delta \tau _{0}\frac{\varphi _{n}}{c^{2}}$.\qquad \qquad \qquad
\qquad \qquad \qquad \qquad \qquad \qquad \qquad \qquad \qquad \qquad \qquad
\qquad \qquad (9)

Thus, we conclude that for appropriate choice of the parameters in (9), we
are able to observe the absolute event: a meeting of two short light pulses
in the spatial point $x_{n}$.

\bigskip

{\bf 2.3. Case of rigid non-inertial frame}

In this Section we will consider the problem in rigid non-inertial frame
moving along the axis $x$ at constant (in relativistic meaning) acceleration 
$a$, and we will perform the exact calculations due to importance of the
results obtained.

By definition, in a rigid frame the proper distance between two spatial
points (measured by means of a scale being at rest in this frame) does not
de-pend on time. Let us define such a rigid frame by the relationships [3]

$x^{\alpha }=x^{\prime \alpha };t^{\prime }=0;t=\tau $,\qquad \qquad \qquad
\qquad \qquad \qquad \qquad \qquad \qquad \qquad \qquad \qquad \qquad \qquad
(10)

where the primer space and time coordinates belong to successive
instantane-ously co-moving inertial reference frames, while $\tau $\ stands
for the proper time at the origin of coordinates. In such definition, for
the case of constant (in instanta-neously co-moving inertial frames)
acceleration $a$ along the axis $x$, a relationship between space-time
coordinates in a fixed inertial reference frame ($T,X,Y,Z$) and ($t,x,y,z$)\
takes the form [3]:

$dT=dt\left( 1+\frac{ax}{c^{2}}\right) \cosh \frac{at}{c}+\frac{dx}{c}\sinh 
\frac{at}{c}$,\qquad \qquad \qquad \qquad \qquad \qquad \qquad \qquad \qquad
\qquad \qquad (11)

$dX=cdt\left( 1+\frac{ax}{c^{2}}\right) \sinh \frac{at}{c}+dx\cosh \frac{at}{%
c}$.\qquad \qquad \qquad \qquad \qquad \qquad \qquad \qquad \qquad \qquad
\qquad (12)

\qquad The metrics of space-time determined by (11), (12), is the following:

$ds^{2}=c^{2}dt^{2}\left( 1+\frac{ax}{c^{2}}\right)
^{2}-dx^{2}-dy^{2}-dz^{2} $.\qquad \qquad \qquad \qquad \qquad \qquad \qquad
\qquad \qquad \qquad (13)

The corresponding components of the metric tensor are:

$g_{00}=\left( 1+\frac{ax}{c^{2}}\right) ;g_{0\alpha
}=0;g_{11}=g_{22}=g_{33}=-1$, all others $g_{\alpha \beta }=0$.\qquad \qquad
\qquad \qquad (14)

($\alpha ,\beta =1...3$). The physical values are defined as

$dx_{ph0}=\sqrt{g_{00}}dx^{0}+\frac{g_{0\alpha }dx^{\alpha }}{\sqrt{g_{00}}}$%
,

$\sum dx_{ph\alpha }^{2}=\left( -g_{\alpha \beta }+\frac{g_{0\alpha
}g_{0\beta }}{g_{00}}\right) dx^{\alpha }dx^{\beta }\qquad \qquad \qquad
\qquad \qquad \qquad \qquad \qquad \qquad \qquad \qquad $(15)

Substituting the components of metric tensor from (14) to (15), we obtain

$dx_{ph}=dx,dy_{ph}=dy,dz_{ph}=dz$;\qquad \qquad \qquad \qquad \qquad \qquad
\qquad \qquad \qquad \qquad \qquad (16)

$dt_{ph}=dt\left( 1+\frac{ax}{c^{2}}\right) $.\qquad \qquad \qquad \qquad
\qquad \qquad \qquad \qquad \qquad \qquad \qquad \qquad \qquad \qquad \qquad
(17)

Due to independence of the metric tensor on time, we again get the equality
(5), from there

$\sum\limits_{m=1}^{n-1}\left( t_{m}^{\prime }-t_{m}\right) =0$.\qquad
\qquad \qquad \qquad \qquad \qquad \qquad \qquad \qquad \qquad \qquad \qquad
\qquad \qquad (18)

\qquad The intervals of physical time at different points are determined by
(17). Hence, the physical values

$\Delta \tau _{0}=\int_{0}^{\Delta t_{m}}dt\left( x_{m}\right) =\Delta
t_{m}\left( 1+\frac{ax_{m}}{c^{2}}\right) $,\qquad \qquad \qquad \qquad
\qquad \qquad \qquad \qquad \qquad \qquad (19)

and

$\Delta t_{m}=\frac{\Delta \tau _{0}}{1+ax_{m}/c^{2}}$.\qquad \qquad \qquad
\qquad \qquad \qquad \qquad \qquad \qquad \qquad \qquad \qquad \qquad \qquad
\qquad (20)

Therefore, $\Delta t_{m}^{\prime }$\ and $\Delta t_{m}$\ are equal to each
other, and the third term in (4) is equal to

$\sum\limits_{m=0}^{n-1}\Delta t_{m}-\sum\limits_{m=1}^{n}\Delta
t_{m}=\left( \Delta \tau _{0}+\sum\limits_{m=1}^{n-1}\Delta t_{m}\right)
-\left( \sum\limits_{m=1}^{n-1}\Delta t_{m}+\Delta t_{n}\right) =$

$=\Delta \tau _{0}-\Delta t_{n}=\Delta \tau _{0}\frac{ax_{n}}{c^{2}}\left( 
\frac{1}{1+ax_{n}/c^{2}}\right) $.\qquad \qquad \qquad \qquad \qquad \qquad
\qquad \qquad \qquad \qquad \qquad (21)

Substituting the obtained values (19), (21) into (4), one gets:

$t_{2}-t_{1}=\Delta t+\Delta \tau _{0}\frac{ax_{n}}{c^{2}}\left( \frac{1}{%
1+ax_{n}/c^{2}}\right) $.\qquad \qquad \qquad \qquad \qquad \qquad \qquad
\qquad \qquad \qquad \qquad (22)

Hence, the left and right light pulses will meet in the point

$x_{n}=-\frac{\Delta tc^{2}}{a\left( \Delta \tau _{0}+\Delta t\right) }$%
.\qquad \qquad \qquad \qquad \qquad \qquad \qquad \qquad \qquad \qquad
\qquad \qquad \qquad \qquad \qquad (23)

\qquad Thus, an observer in an accelerated frame will detect the absolute
event: the left and right light pulses will meet in the point defined by
(23) (under negative sign of the acceleration $a$). This conclusion is in
agreement with the result of Section 2.2 and the equivalence principle.
However, here we meet a quite difficult problem: for observer in external
inertial frame both light pulses will never intersect.

\qquad Indeed, let the process of light pulses propagation in the
accelerated frame be observed from some inertial reference frame.
Furthermore, let us choose for observing the light pulses propagation
process an inertial frame K, such that at the time moment when an observer
sees the appearance of the left light pulse in the point $x=0$, he
simultaneously sees an arriving right pulse to RL$_{1}$ (such a choice is
always possible due to (1)). For this time moment, let us introduce into
consideration the second inertial frame K$_{s}$ shifted along the axis $x$
at such a distance (with respect to K) which is equal to the distance
between RL$_{0}$ and RL$_{1}$. (The relative velocity of K and K$_{s}$ is
equal to zero). Due to the space homogeneity in inertial frames, such a
shift is equivalent to re-numeration of the re-emitters in K$_{s}$: the RL$%
_{m}$ (in K) be RL$_{m-1}$ (in K$_{s}$). Hence, the propagation time from RL$%
_{0}$ to RL$_{n-1}$ for the left pulse is exactly equal to the propagation
time from RL$_{1}$ to RL$_{n}$ for the right pulse in both K and K$_{s}$
frames (since the RL$_{1}$, RL$_{n}$ in K are the RL$_{0}$, RL$_{n-1}$ in K$%
_{s}$). Hence, at the moment of time (in K) when the right pulse is emitted
by RL$_{n}$, the left one is emitted by RL$_{n-1}$ for any $_{n}$.
Therefore, the light pulses considered will never meet in the inertial frame
K, that means a contradiction with the causality principle.

{\bf Conclusions}

\qquad Thus, a consistent relativistic consideration of the problem about
special kind of two light pulses propagation in a non-inertial reference
frame contradicts with the causality principle. One can show that a
resolution of this contradiction is possible only under supposition that a
geometry of empty space-time is not pseudo-Euclidean [4]. However, it is in
a deep disagreement with the relativity theory. It seems that a full
resolution of this contradiction within the scope of relativity theory is
impossible.

\bigskip

{\bf Bibliography}

[1] \qquad Logunov, A.A. (1997) - Lectures on the Theory of Relativity and
Gravi-tation (modern analysis of the problem). Nauka. Moscow (in Russian).

[2]\qquad Kholmetskii, A.L. (1997) -- On ``General kinematics'' in an empty
space. Physica Scripta. 55, 18.

[3]\qquad Moller, C. (1972) - Theory of Relativity. Clarendon Press. Oxford.

[4]\qquad Kholmetskii, A.L. (1997) -- Extended General kinematics of an
empty space. Physica Scripta. 56, 539.

\end{document}